# Leveraging Internet of Things Network Metadata for Cost-Effective Automatic Smart Building Visualization

Benjamin Eichler Staugaard[1][0009-0007-3392-4567], Simon Soele Madsen[1][0009-0008-5279-9500], Zheng Ma[1][0000-0002-9134-1032], Salman Yussof[2][0000-0002-2040-4454], and Bo Nørregaard Jørgensen[1][0000-0001-5678-6602]

[1] SDU Center for Energy Informatics, Maersk Mc-Kinney Moller Institute, University of Southern Denmark, Odense, Denmark
[2] Institute of Informatics and Computing in Energy, Universiti Tenaga Nasional, Kajang 43000, Malaysia
beis@mmmi.sdu.dk, ssma@mmmi.sdu.dk, zma@mmmi.sdu.dk, salman@uniten.edu.my, bnj@mmmi.sdu.dk

**Abstract.** In recent years, the building sector has experienced an increasing legislative pressure to reduce the energy consumption. This has created a global need for affordable building management systems (BMS) in areas such as lighting-, temperature-, air quality monitoring and control. BMS uses 2D and 3D building representations to visualize various aspects of building operations. Today the creation of these visual building representations relies on labor-intensive and costly computer-aided design (CAD) processes. Hence, to create affordable BMS there is an urgent need to develop methods for cost-effective automatic creation of visual building representations. This paper introduces an automatic, metadata-driven method for constructing building visualizations using metadata from existing smart building infrastructure. The method presented in this study utilizes a Velocity Verlet integration-based physics particle simulation that uses metadata to define the force dynamics within the simulation. This process generates an abstract point cloud representing the organization of BMS components into building zones. The developed system was tested in two buildings of respectively 2,560 m² and 18,000 m². The method successfully produced visual building representations based on the available metadata, demonstrating its feasibility and cost-effectiveness.

## 1    Introduction

In recent years, there has been a significant increase in focus on smart buildings, green building technology, and building energy performance. This trend is driven by rising legislative pressure and directives that require buildings to comply with new requirements for automatic lighting, temperature control, and indoor climate monitoring [1].

Buildings are expected to meet these standards within the near future. However, many older buildings do not meet these new requirements.

To address this issue, buildings must be retrofitted with hardware that allows them to meet the new requirements. This involves installing sensors and actuators and implementing software solutions for managing and monitoring these systems. In most cases, the most cost-effective retrofitting strategy is wireless retrofitting using Internet of Things (IoT) devices, i.e., sensors and gateways, as it does not require expensive invasive changes such as installing new cabling within the building's walls [2]. This shift necessitates state-of-the-art, cost-effective tools for monitoring and maintenance that can intelligently manage the installed smart infrastructure.

An important aspect of such tooling is how the installation is communicated and utilized by various stakeholders in an efficient and understandable manner. The optimal visual representation of a building is arguably a three-dimensional representation that any stakeholder can immediately comprehend, as it closely resembles the physical structure. While computer-aided design (CAD)-based building visualizations are widely used for this purpose, a significant drawback is their labor-intensive and costly nature. These visualizations require professionals to develop them from building plans, which are often outdated, incomplete, or missing, particularly in older buildings [3].

This makes them unsuitable to adopt for economically challenged building owners. However, even in this case, having a 3D visualization as a communication medium offers significant value. It enhances facility management, streamlines maintenance efforts and planning, aids future renovations and retrofitting, and ultimately leads to cost savings and improved energy efficiency [3].

To make building visualization affordable to any building retrofitted with an IoT infrastructure at minimal to no cost, this paper proposes a method for automatic generation of 3D building visualizations from the metadata already available in the building management system (BMS). The metadata describes the installed IoT network encompassing IoT sensors and gateways, their connectivity, and how these are allocated in the building. Using this proposed method, it is possible to eliminate the need for expensive manual visualization creation.

The proposed method aims to efficiently produce visualizations using a Velocity Verlet physics particle simulation that produces an initial point cloud based on the metadata from the BMS [4]. This method allows force dynamics to be applied to the particles, effectively driving the layout through attractive and repulsive forces. At each timestep in the simulation, an arbitrary number of metadata is used to determine the forces between nodes. Finally, the simulation decorates the model by encapsulating grouped BMS components such as IoT sensors and gateways into rooms, floors, and buildings, finalizing the visualization. This method is chosen because it relies solely on metadata, making it essentially labor-free to apply. Additionally, it allows the visualization to dynamically adapt over time as the IoT installation evolves and the metadata change.

To validate the functionality of the method, two case studies were conducted. One took place at the University of Southern Denmark in Odense, Denmark, and the other at Universiti Tenaga Nasional (UNITEN) in Malaysia. The buildings were 2,560 square meters and 18,000 square meters in size, respectively. These case studies aimed to validate the method's visualization capabilities and quality.

The paper is organized as follows: Initially, a literature review is conducted to reveal the current state of the art in building visualization strategies and methodologies. Following this, the methodologies applied in this paper are presented and justified. Subsequently, the case studies and their contributions are discussed. The results are then presented, demonstrating the application of the physics particle simulation paired with metadata and transformation strategies to construct a building visualization. Finally, the results are discussed, and conclusions are drawn, with suggestions for future research and improvements.

## 2 Literature Review

The literature review focuses on the intersection of smart buildings, the Internet of Things (IoT), simulation, and software development, with a particular emphasis on visualization rendering software and technologies. The aim is to identify approaches for data-driven building generation and visualization, with a specific interest in IoT data-driven building visualizations. Several strategies for rendering and visualization have been explored in the literature. While few studies share the exact goal of this paper, the technologies utilized exhibit commonalities that are relevant regardless of the specific objectives. These technologies can be broadly categorized as follows.

**Computer-Aided Design (CAD) Software** is utilized for precise engineering and architectural design, enabling the creation and modification of detailed 2D and 3D models. Notable examples include AutoCAD and Autodesk Revit by Autodesk. Autodesk has been experimenting with generative design for CAD-based models. However, this approach still requires significant manual effort, maintaining the human-in-the-loop aspect and necessitating manual adjustments [5]. Other studies have explored various applications of CAD software in smart building design, including the integration of environmental data for more sustainable architectural solutions and the use of parametric design to optimize space utilization [6, 7].

**Web Graphics Library (WebGL)** is a JavaScript API for rendering 3D graphics in web browsers without plugins, enabling real-time visualization and interaction. It is widely used in web-based applications. Technologies like Three.js leverage WebGL to make 3D models more accessible through web interfaces, though these still often rely on pre-existing models converted from various 3D formats [8-10].

**Augmented Reality/Virtual Reality (AR/VR) Technologies** create immersive experiences by overlaying digital content onto the real world (AR) or simulating entire environments (VR). They are used in various applications, including architectural visualization and virtual production. Studies have explored the integration of IoT sensor data with Building Information Models (BIM) in AR contexts to visualize live environmental data. However, these do not utilize the data for generating the models themselves [11-13].

**Game Engines** such as Unity [14] and Unreal Engine [15] provide frameworks for developing video games, offering tools for rendering, physics simulation, and scripting.

These tools are also applied to architectural visualization, allowing for dynamic and interactive model generation [16]. Most existing technologies share a common limitation, they rely on static and fixed models, schematics, or blueprints. Specifically, AR, and VR often depend on loading external CAD files, which require significant manual effort to create and maintain. Although WebGL and game engines are more suitable for generating dynamic models, they too generally start from predefined data. For instance, a study investigated integrating IoT sensor data with CAD-based models for AR applications, using sensor data to visualize live environmental information in a building context. However, the model generation still required manual intervention and did not dynamically adapt to changes in the building structure [11]. Another study proposed an automatically generated model from spatial data collected by an autonomous robot. This approach created a point cloud to be processed into a building model, but it remained labor-intensive and static, necessitating extensive effort to update as the physical building changed [6].

Hybrid approaches also exist, where CAD-based file formats are converted for use in WebGL contexts. Using file conversion plugins, CAD models can be translated into formats suitable for web-based rendering technologies like Three.js [17], making the models more accessible through web browsers [8].

The literature includes numerous methods for building visualization and rendering, yet there is a clear lack of fully automated generation methods. Existing methods rely on fixed starting points, such as schematics or building plans, which are costly to produce and maintain. As buildings evolve, these models quickly become outdated unless significant resources are devoted to updating them.

## 3    Methodology

The selected rendering technology is web-based, specifically utilizing WebGL and Three.js. As identified in the literature review, WebGL is highly suitable for real-time visualization and interaction within web browsers, making the generated models widely accessible and easy to distribute. Web-based rendering technologies such as Three.js allow for seamless integration of dynamic models, offering significant advantages over traditional CAD software, which often requires extensive manual effort and relies on static models [8-10]. This approach leverages the strengths of WebGL to provide an interactive and user-friendly visualization platform, ensuring accessibility and broad usability.

The selected particle simulation is based on the Velocity Verlet integration algorithm [4]. This algorithm is chosen due to its stability and efficiency in simulating the motion of particles under various forces. In this context, the particles represent IoT devices in the building's IoT network, and their interactions are driven by the IoT network metadata. Unlike traditional visualizations that depend on predefined structures, this approach allows the simulation to be highly adaptive, responding dynamically to changes in the metadata. This algorithm is particularly suited for this purpose as it provides results with low computational cost and no manual effort. This approach is not seen in any of the investigated literature.

The metadata driving the simulation is extracted from the IoT network within the building. This metadata is queried, made available to the simulation, and translated into forces that influence the positioning of particles within the simulation.

1. **Querying Metadata**: The system continuously queries the IoT network for updated metadata, ensuring the simulation reflects the most current layout and configuration of the IoT network in the building.
2. **Translating Metadata into Forces**: The metadata is translated into forces that the simulation understands. For example, sensor allocation data helps to define the spatial distribution of sensors, signal strength can indicate proximity and connectivity between different areas, and relative room positions provide a basis for organizing the spatial structure of the building. Investigated BMS metadata strategies can be seen in Table 1.
3. **Iterative Process**: The simulation runs iteratively, applying the translated forces to the particles. An alpha value is used to control the convergence of the simulation. A larger alpha value allows the simulation more iterations to position particles at the cost of more computation power and vice versa. This iterative process continues until the alpha value decreases to zero, indicating that the visualization is finalized.

This methodology currently focuses on spatial and connectivity data from the IoT network. Indoor climate data, such as temperature and humidity, which are readily available from IoT sensors, can be easily integrated to enrich the generated model. This could provide a more comprehensive representation of the building environment.

**Table 1.** BMS metadata strategies

| Strategy | Description |
| --- | --- |
| Resource allocation | The resource allocation strategy is the most critical metadata source. It describes the rooms in the building and the devices' association with these rooms. This information is used to group IoT devices and associates their positions with specific floors and rooms within the building. |
| Signal strength | In a wireless IoT network setup, signal strength can be extracted from installed IoT sensors at any given time. This data is used to create edges between sensors and gateways in the simulation, where edge distance is proportional to signal strength. Poor signal strength indicates a greater distance from the gateway, while strong signal strength suggests proximity. |
| Infrastructure materials | Signals weakened by obstacles such as walls can be accounted for by incorporating metadata about building materials. Different materials attenuate the signal to varying degrees. By applying a weakening factor based on the material, the simulation can adjust the link distances to reflect more accurate physical proximities. |
| Relative rooms | Resource allocation groups IoT devices within rooms, but to position rooms accurately, metadata about their relative positions is needed. This can be collected during installation. Relative positions can sometimes be inferred from device |
| | links, where connections across rooms suggest closer proximity. Signal strength can further refine these inferred positions. |
| Room size | Room size can be determined in two ways: by computing it from the IoT devices within the room if no size information is available or by using known room dimensions. When computing from IoT devices, the link distances from sensors to gateways can be used to create a bounding box on a 2D plane, providing an inferred room size based on resource allocation and signal strength. |
| Room shape | To finalize the building visualization, rooms must be decorated with geometries that reflect their actual shapes. For non-rectangular rooms, feeding geometry information into the simulation allows it to construct complex room shapes and accurately position them within the building model. |
| Schematic parsing | Building schematics can simplify the process, potentially rendering other metadata redundant. However, schematics are static and may be incorrect or incomplete. Introducing metadata representing schematic knowledge is conceptually possible but requires careful validation against dynamic data. |
| Geographical Information | If collected during the installation, Geographic Information System (GIS) data can be used to create reference points with spatial coordinates [10, 18]. This can enhance IoT device positioning accuracy and provide a framework for integrating additional graphical information. |

Each of these strategies complement one another, enhancing the accuracy of the building visualization. The only required strategy is the resource allocation strategy which produces an abstract representation. Other strategies improve the fidelity of the visualization. The process flow of the proposed method for generating building visualizations can be seen in Fig. 1. The process is incremental and iterative. It starts with an initial random configuration and continuously adapts based on the metadata.

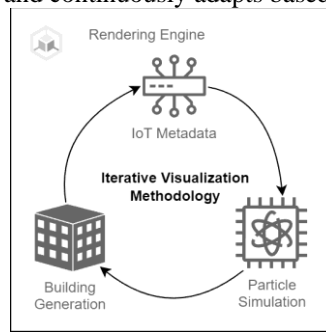

**Fig. 1.** The iterative visualization methodology applied in the rendering engine.

By leveraging IoT network metadata and advanced simulation techniques, this methodology offers a robust and dynamic approach to smart building visualization. The integration of web-based rendering technologies ensures that the resulting models are not only accurate and up to date but also easily accessible and interactive for users.

## 4     Case Studies

The effectiveness and quality of the proposed method for generating building visualizations have been verified using two case study buildings, one located at the University of Southern Denmark seen in Fig. 2 and another at Universiti Tenaga Nasional in Malaysia seen in Fig. 3. Each case study building was retrofitted with an IoT network where each IoT device were associated with a room ID. The room ID is represented by a QR code as explained in [19]. The first case study building is the Maersk Mc Kinney Moller Institute at the University of Southern Denmark. It has two floors and a total floor area of 2,560 m2. Currently the IoT network covers the upper floor with 17 registered rooms, 88 IoT sensors and 5 IoT gateways installed.

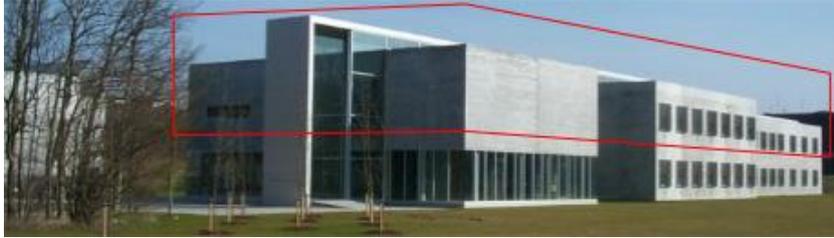

**Fig. 2.** The Maersk Mc Kinney Moller Institute at the University of Southern Denmark.

The second case study is the library building at Universiti Tenaga Nasional in Malaysia. It has a total floor area of 18,000 m2. Currently the IoT network covers the 5 floors of the main building with 57 registered rooms and 212 sensors and 21 gateways installed.

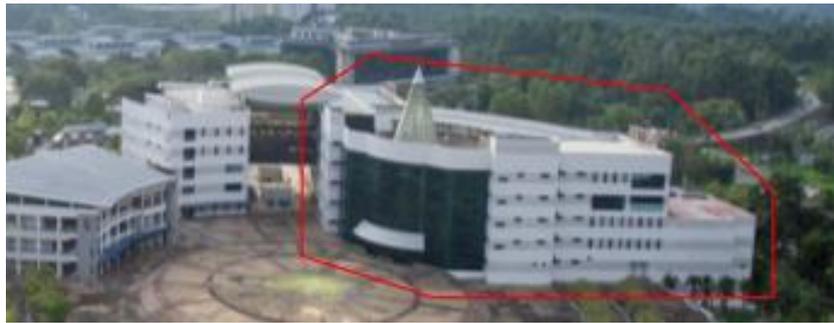

**Fig. 3.** The library building at Universiti Tenaga Nasional in Malaysia.

The simulation executes its layout strategies in continuous steps, incrementally positioning the particles based on the IoT network metadata. The results of the case studies are presented in the following section.

## 5      Results

The results section follows a chronological order, presenting the core implementation of the physics engine, the utilization of metadata to drive forces in the particle simulation, and the subsequent transformation of the simulation into a visual building representation. The data used in the figures are test data intended to demonstrate the contribution of each step. Additionally, the results of the individual case studies conducted at the University of Southern Denmark and Universiti Tenaga Nasional utilizing real data from the physical installations are presented.

### 5.1    Physics Particle Simulation Realization using Velocity Verlet Numerical Integration

The proposed method for generating the metadata-based building visualization is using a physics particle simulation, as described in the Methodology section. A flowchart of the algorithm for running the simulation can be seen in Fig. 4.

Initially, metadata is extracted from the installed IoT network. The metadata defines the forces between the particles, thereby directly influencing the velocities of the particles, and hence their relative position to each other. To improve the visual quality of the results, collision detection ensures that no particles intersect, with corrective measures implemented if necessary.

Throughout the simulation, the alpha value gradually decreases over time, symbolizing the cooling down of the system. Once the alpha value reaches zero, the simulation concludes, resulting in particle positions that are based on the input metadata.

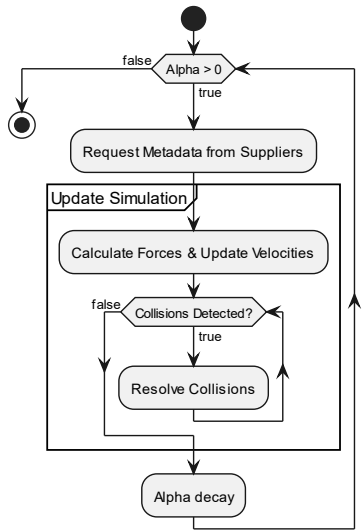

**Fig. 4.** Particle simulation pseudo-algorithm.

Leveraging IoT Metadata for Cost-Effective Automatic Building Visualization 9

The outcome is an abstract point cloud, where each point is a specific node from Fig. 5. Each node serves various purposes within the simulation, as described in Table 2.

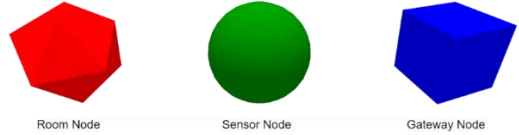

**Fig. 5.** Simulation node types.

**Table 2.** Table of node descriptions.

| Node | Purpose |
| --- | --- |
| Room Node | The room node serves as a central reference point for all sensor and gateway nodes associated with a specific room. This simplifies the grouping process for subsequent steps in the simulation. However, it's important to note that there are limitations associated with this approach, which are discussed in further detail in the Discussion section. |
| Sensor Node | Sensor nodes represent individual IoT sensors within the installed IoT network. They are typically connected to a gateway node, with the link distance between the sensor node and the gateway node determined by metadata. In wireless installations, signal strength metadata is often utilized, potentially augmented by building construction metadata regarding signal penetration through walls. |
| Gateway Node | Gateway nodes represent the IoT gateways within the system to which all IoT sensors are connected. These nodes play a crucial role in maintaining the spatial coherence of the simulation. By linking gateway nodes to room nodes, the simulation ensures that IoT sensor nodes remain grouped within their respective rooms, facilitating overall spatial organization. |

The following simulation encompasses 10 rooms distributed across three floors, featuring a gateway situated on each floor, and five sensors allocated to each room. Prior to applying any forces to the simulation, the algorithm generates a simulation outcome depicted in Fig. 6. This initial outcome serves as the foundation for constructing the building before executing the next steps.

The initial simulation result manifests as an abstract point cloud in disarray, with nodes positioned randomly and lacking mutual awareness. However, the subsequent sections delve into strategies employed to rectify this random arrangement, elucidating how the simulation transitions from its initial state to a coherent organization resembling a visual representation of the building.

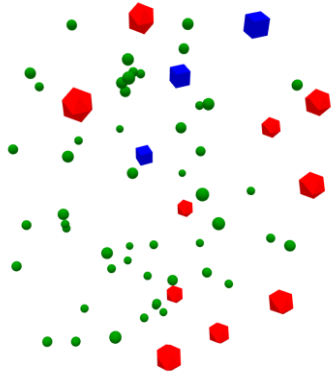

**Fig. 6.** Simulation result without forces.

Now, by adding IoT network metadata, it is possible to drive force dynamics within the particle simulation, thereby guiding the nodes toward structured positions. The forces exerted on the nodes are outlined in Table 3.

Translating the metadata into forces the simulation understands is critical. While link forces statically space two nodes, strong forces can alter a link's length. Additionally, forces propagate through links, so changes in one node's position affect linked nodes.

A fundamental force within the simulation is charge, where nodes can possess either a negative or positive charge, resulting in repulsion or attraction towards other nodes, respectively. This concept underscores the intricate interplay of forces crucial for the simulation's operation, emphasizing the necessity of accurately translating metadata into actionable forces.

**Table 3.** Simulation forces between nodes.

| Relation | Force | Influence |
| --- | --- | --- |
| Sensor to Gateway | Attract | Establishes an attractive force between a sensor and its associated gateway, ensuring their cohesion within the simulation. This force is determined based on metadata such as signal strength, enabling nodes to be positioned at appropriate distances relative to one another. |
| Gateway to Room | Attract | A gateway is associated with a given room at any time. As gateways are consistently linked to specific rooms, this force reinforces their proximity, enhancing the simulation's fidelity. |
| Sensor to Room | Attract | Imposes an attractive force from sensors towards their respective rooms, creating cohesive grouping within the simulation. Although sensors are inherently linked to gateways, this force can be essential in scenarios where sensor-gateway |
| | | connections are not established or where such information is lacking. |
| Device to Device | Both | Generates an attraction force between IoT devices within the same room while concurrently exerting repulsive forces between devices in distinct rooms. This dual-action mechanism ensures that devices remain appropriately spaced within their respective rooms. |
| Room to Room | Repulse | A repulsive force between rooms, preventing overlapping and maintaining spatial separation. By mitigating room overlap, this force contributes to the refinement of the simulation's spatial representation and quality. |
| Floor to Floor | Repulse | Provides an optional repulsive force between floors, serving to maintain vertical separation within the simulation. While the y-coordinates of nodes are typically fixed based on floor level, the inclusion of this force offers additional control over floor-to-floor spacing. |

Applying designated forces to the simulation repositions nodes based on metadata. Nodes are anchored to fixed y-coordinates for each building floor, and sensor nodes are grouped by room. Each sensor node links to a gateway node. Initially, links are uniform in length, but signal strength data introduces variability, affecting link lengths. The resulting simulation can be seen in Fig. 7.

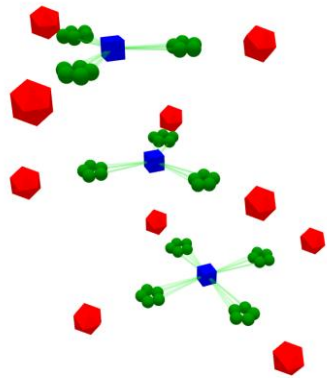

**Fig. 7.** Simulation with forces.

Fig. 7 depicts the simulation advancing towards an abstract metadata-driven layout, albeit lacking the definitive characteristics of a building structure. Various methodologies can be employed to bridge this gap, each offering different levels of complexity. To simplify the process, the current strategy involves enveloping the simulation with axis-aligned bounding boxes (AABB). A pseudo algorithm, outlined in Fig. 8, orchestrates this operation by collecting device data, computing mesh points, calculating cartesian extremes, and expanding the AABB for enhanced visual representation. Subsequently, the simulation is enhanced with room meshes, a step that can be extended to encompass the entire building, with further stylization possibilities.

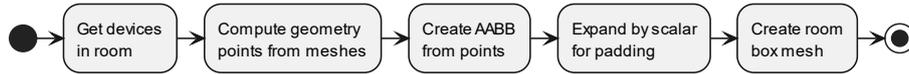

**Fig. 8.** Simulation skinning method.

Extending this algorithm facilitates the creation of an aggregate AABB encapsulating the entire building structure. Enhancing this with 2D planes to separate different floors yield the results depicted in Fig. 9. Despite limited metadata availability, the simulation proficiently generates a valid representation based solely on the presence and locations of IoT devices. Enhanced spatial fidelity can be achieved through the provision of additional metadata inputs to the simulation algorithm.

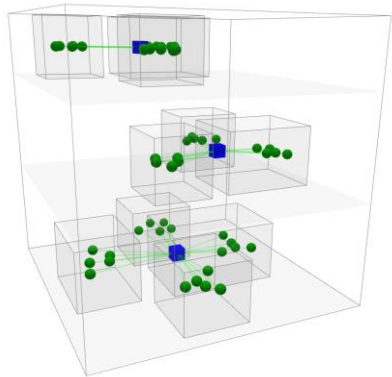

**Fig. 9.** Skinned building simulation.

### 5.2   University of Southern Denmark

The University of Southern Denmark served as one of the case studies, where the aim was to construct a visualization for its IoT network. Primarily deployed to monitor and regulate indoor climate conditions, the IoT devices were mapped to specific rooms within the building using QR codes during installation. The available metadata for the case study comprised resource allocation and signal strength data.

Leveraging this dataset, the simulation generated the building visualization seen in Fig. 10. The visualization incorporated dynamically colored links reflecting signal strength. The inclusion of room labels provided additional context. The simulation algorithm intelligently inferred relative room positions based on cross-room connections, enhancing spatial accuracy. While the model's fidelity could be further refined with additional room layout and relative positioning metadata, the current iteration stands as a satisfactory representation, laying the foundation for future development.

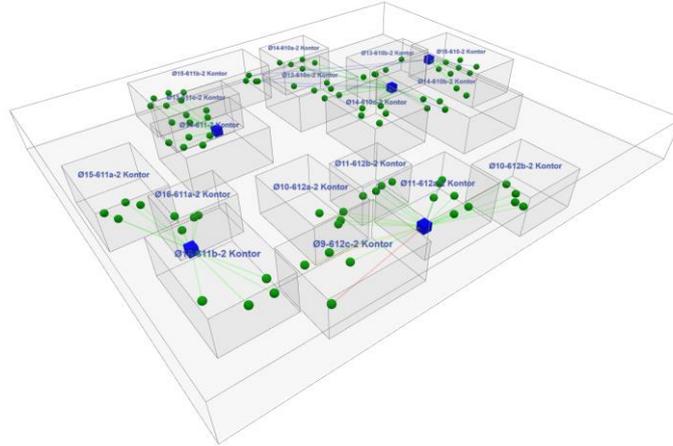

**Fig. 10.** University of Southern Denmark building simulation.

### 5.3   Universiti Tenaga Nasional (UNITEN)

The second case study conducted at Universiti Tenaga Nasional presented a unique opportunity to assess the scalability and robustness of the simulation framework in a larger building. Distinguished by its considerable size in comparison to the previous case study, this building offered an ideal environment to evaluate the simulation's performance with a larger dataset. The resulting visualization, depicted in Fig. 11, underscores the simulation's capability to efficiently process metadata and generate a comprehensive virtual representation instantaneously.

However, the increased scale of the visualization introduced challenges associated with spatial occlusion, primarily manifested through overlapping geometries. While this issue is predominantly pertinent to static images, the simulation mitigated its impact by facilitating dynamic interaction. By enabling users to manipulate the camera perspective, toggle geometries, adjust floor views, and engage with various interactive features, the simulation experience was markedly enhanced.

Overall, the case study at Universiti Tenaga Nasional underscored the simulation's adaptability to diverse building contexts and its capacity to effectively manage larger datasets while offering an intuitive user experience.

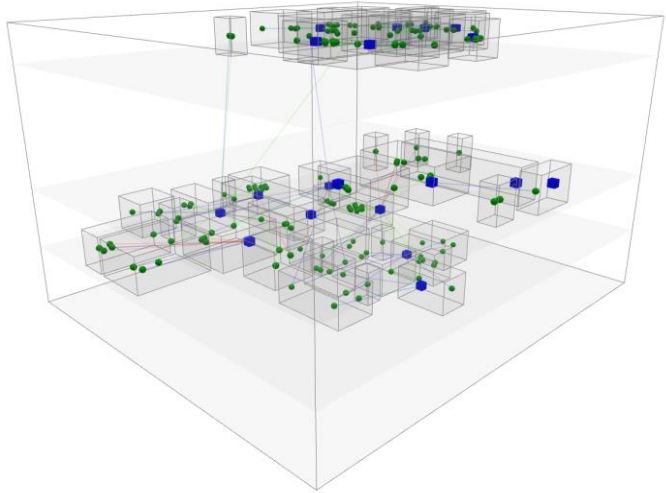

**Fig. 11.** Universiti Tenaga Nasional building simulation.

## 6    Discussion

The core objective of this paper was to develop a method for generating 3D building visualizations using IoT network metadata, employing a particle simulation to create a force-directed graph enhanced with bounding-boxes to resemble the building.

When compared to the state of the art, existing methods rely heavily on fixed schematics or building plans, which result in highly accurate visual representations. However, these traditional methods are costly and inflexible, particularly when buildings undergo changes. The proposed method offers a flexible and cost-effective alternative, dynamically adapting to changes in the metadata. Despite its current experimental nature and the need for refinement to enhance visual accuracy, this method provides a promising foundation for future development.

The initial challenge of generating a building visualization from metadata has been successfully addressed in this study. The algorithm demonstrated its capability to produce building visualizations in line with the quality of the provided metadata. However, the visual fidelity often diverged from the physical buildings due to the limitations inherent in the metadata from the case studies.

The lower visual accuracy observed in the case studies when comparing to their physical counterparts underscores the unique challenges of this method. While the cost-effective, dynamic, and automatic approach presents a significant advantage over traditional methods, the shortcoming lies in the visual fidelity. At this stage in the research, the focus was on developing a functional method rather than achieving perfect visual fidelity. Consequently, the divergence between generated models and physical buildings has not been quantified, as the priority was to establish a working method. Quantifying this divergence is planned for future work as the method is refined and the visual fidelity improves.

Future work will aim to reduce the gap between the generated visualizations and their physical counterparts. An ongoing effort involves incorporating GPS coordinate metadata during IoT network installation. This could enhance visual fidelity by providing more precise spatial information, leading to more accurate and realistic building visualizations in the future.

## 7    Conclusion

This paper sought to address the challenges associated with the manual creation of 3D building visualizations, particularly their high costs and labor-intensive nature. Through the introduction of a metadata-driven approach, this study proposes a solution that aims to broaden access to 3D building visualizations, making them more readily available and cost-effective for stakeholders.

The methodology utilizes efficient rendering software alongside physics simulations to generate building visualizations from metadata extracted from IoT networks. The fidelity and precision of these visualizations directly correlate with the quality of the metadata, emphasizing the crucial role of robust metadata acquisition processes.

Distinguishing itself from traditional methods reliant on fixed building information, this solution offers dynamic adaptability and scalability, capable of evolving alongside changing building environments without incurring substantial additional expenses. This dynamic nature ensures that the visualizations remain relevant and useful over time, accommodating evolving needs and requirements.

Moreover, this methodology presents opportunities for integration into existing IoT ecosystems, enhancing their capabilities with advanced visualization functionalities at minimal overhead. While the current framework operates independently of metadata frameworks or ontologies, future research could explore their integration to further enhance interoperability and facilitate seamless data exchange.

Expanding the scope of metadata strategies beyond those explored in this paper holds promise for enriching the depth and detail of the generated visualizations. Additionally, incorporating user interaction features and mechanisms for persisting modifications could enhance the utility and practicality of the visualizations, enabling stakeholders to actively engage with and manipulate the visualizations as needed.

In conclusion, this study establishes a new approach to building visualization, providing a flexible and cost-efficient solution that offers valuable insights into building environments, enhancing facility management. With further research and innovation, metadata-driven visualization has the potential to bring significant benefits, enhancing efficiency and supporting informed decision-making.

**Acknowledgments.** This paper is part of the project "CELSIUS - Cost-effective large-scale IoT solutions for energy efficient medium- and large-sized buildings", funded by the Danish funding agency, the Danish Energy Technology Development and Demonstration (EUDP) program, Denmark (Case no. 64020-2108)

**Disclosure of Interests.** The authors have no competing interests to declare that are relevant to the content of this article.